\documentclass[preprint,showpacs,preprintnumbers,amsmath,amssymb]{revtex4}
\usepackage{graphicx,epstopdf}

\begin{document}
\title{Acoustic charge  transport in n-i-n three terminal device}
\author{Marco Cecchini}
\email{m.cecchini@sns.it}
\author{Giorgio De Simoni}
\author{Vincenzo Piazza}
\author{Fabio Beltram}
\affiliation{NEST-INFM and Scuola Normale Superiore, I-56126 Pisa,
Italy}
\author{H. E. Beere}
\author{D. A. Ritchie}
\affiliation{Cavendish Laboratory, University of Cambridge,
Cambridge CB3 0HE, United Kingdom}


\begin{abstract}
We present an unconventional approach to realize acoustic charge transport 
devices that takes advantage from an original input region geometry in 
place of standard Ohmic input contacts. Our scheme is based on a n-i-n 
lateral junction as electron injector, an etched intrinsic channel, a 
standard Ohmic output contact and a pair of in-plane gates. We show 
that surface acoustic waves are able to pick up electrons 
from a current flowing through the n-i-n junction and steer them toward 
the output contact. Acoustic charge transport was studied as a function 
of the injector current and bias, the SAW power and at various temperatures. 
The possibility to modulate the acoustoelectric current by means of 
lateral in-plane gates is also discussed.
The main advantage of our approach relies on the possibility to drive 
the n-i-n injector by means of both voltage or current sources, thus 
allowing to sample and process voltage and current signals as well.
\end{abstract}

\maketitle

The introduction of acoustic charge transport (ACT) devices\cite{hoskins-apl82} 
brought true digital programmability to analog signal
processing.  ACT devices are wideband, operating up to several 
GHz and can have digital programmability of hundreds of individual 
taps\cite{guediri-ieee87}. Moreover, these devices are based on compound semiconductor
materials (i.e. GaAs) and can be easily integrated with the 
existing technology.

ACT devices are based on the transport of charge in a piezoelectric 
semiconductor by means of surface acoustic waves (SAWs)\cite{edjeou-sse00}.
Lattice deformations induced by SAWs in a piezoelectric substrate 
are accompanied by potential waves which can trap
electrons in their minima and drag them along the SAW propagation 
direction resulting in a net dc current or voltage
(the acoustoelectric effect)\cite{wixforth-surfsci94,wixforth-ssc92,campbell-ssc92}. In ACT devices the SAW electric 
field bunches electrons together into packets and
transports them through a semiconductor channel. Typically, this is 
depleted from charge by top and/or back gate
electrodes, while electron packets are extracted from an undepleted 
region of the semiconductor located beneath an
ohmic input contact (IC)\cite{MillerBook,ratolojanahary-iee04}. A time-varying electrical signal 
applied to the IC produces a sequence of charge packets
that travel through the device toward an ohmic output contact 
(OC). The amount of charge in each packet varies depending on the input signal intensity, making this sequence represent a sample ready for digital processing. The 
described ACT device adds a time delay to the input signal,
depending on the length of the channel. Top 
metallic electrodes along the SAW path allows to modify
the distribution of charge in the packets and to process
the input signal.

In this letter we introduce an alternative approach to realize 
ACT devices, based on an original input region geometry.
The device consists of: i) a n-i-n lateral junction as an electron 
injector; ii) an etched intrinsic channel; iii) a pair of in-plane
gates; iv) a standard Ohmic OC. As we shall show, SAWs can collect electrons from a current 
flowing through the n-i-n junction and steer them toward the 
OC across the intrinsic channel. The main 
advantage of this geometry is the possibility to drive the
n-i-n injector by means of both voltage or current, thus 
allowing to sample and process voltage or current signals as
well.

\begin{figure}[!ht]
\includegraphics[width=8cm]{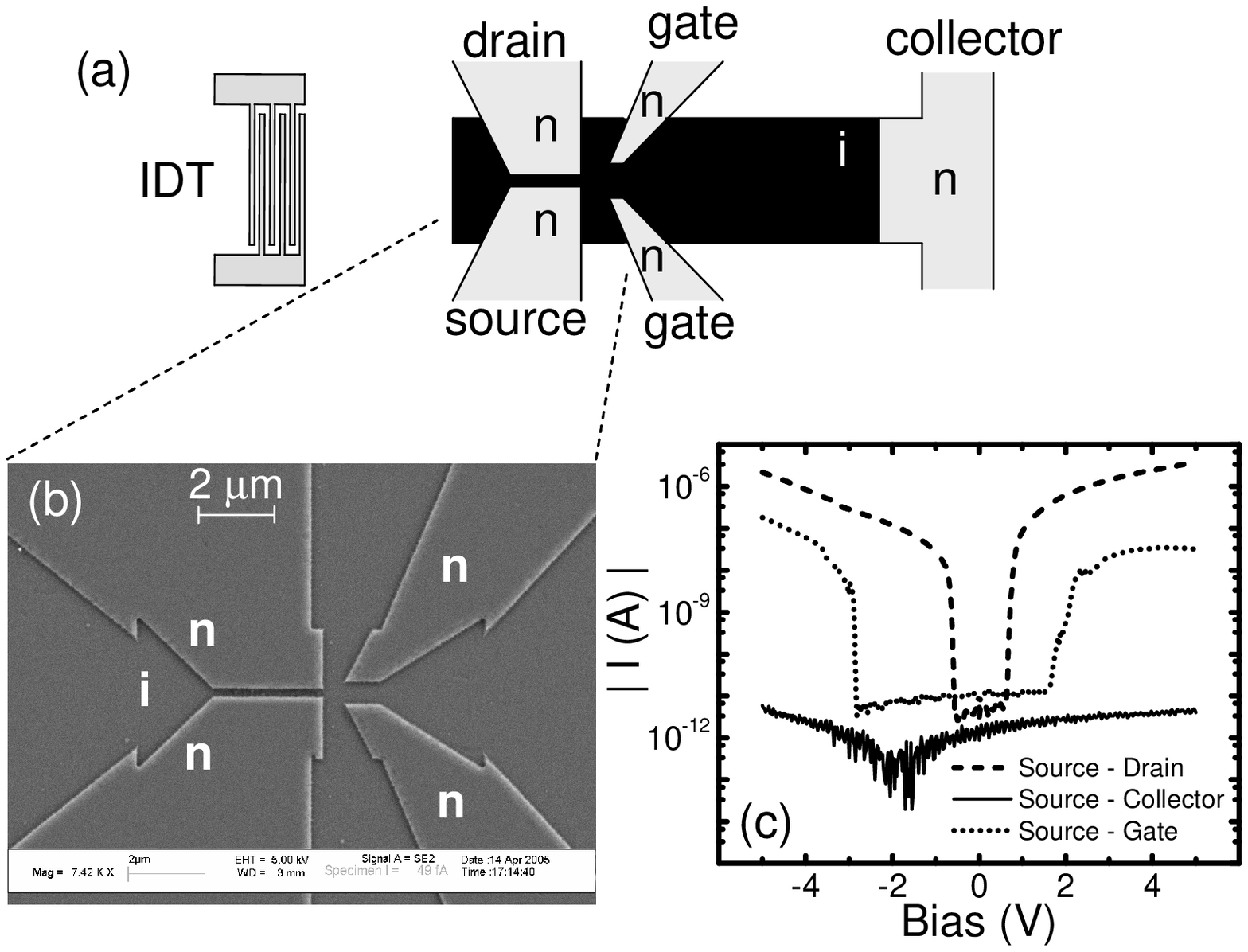}
\caption{(a) Schematic view of the device. (b) Scanning electron 
microscope image of the n-i-n injector region. (b) Source-drain 
(dashed line), source-collector (solid line) and source-gate 
(dotted line) current-voltage characteristics.}
\label{fig1}
\end{figure}

The device was fabricated starting from a n-type modulation-doped 
Al$_{0.3}$Ga$_{0.7}$As/GaAs heterostructure grown by
molecular-beam epitaxy, containing a two dimensional electron 
gas (2DEG) within a 30-nm-wide GaAs quantum well embedded
90 nm below the surface. The measured electron density and 
mobility after illumination at 1.5 K were
$3.33\times10^{11}$ cm$^{-2}$ and $2.10\times10^{6}$ cm$^{2}$/Vs, 
respectively. The heterostructure was processed into
mesas with annealed n-type Ni/AuGe/Ni/Au 
(10/180/10/100 nm) Ohmic contacts by standard optical
lithography, wet chemical etching and thermal evaporation. 
The n-i-n lateral-junction was fabricated according to the
scheme of Fig. \ref{fig1} (a).
Two portions of the 2DEG (named ``source'' and ``drain'') were 
separated by a thin intrinsic spacer (which we shall
call ``the barrier'' in the following) defined by
electron beam lithography and shallow ($\sim 30$ nm) 
etching of the surface. We observed that a 30-nm etching
leads to QW depletion allowing to create intrinsic regions
within the mesa. A 70-$\mu$m long intrinsic region (``the channel'' 
in the following) defined by the same etching step described above 
separates the n-i-n region from a third electron reservoir 
(the ``collector''.)

A pair of lateral control gates
consisting of portions of 2DEG were also fabricated to allow further
control on the collector current
\cite{gloos-prb04}. Figure \ref{fig1} (b) displays a 
scanning electron microscope image of the injector and gate region.

The source-drain separation was chosen of 250 nm in order to have
a breakdown voltage of approximately 1 V. Figure \ref{fig1}(c) shows the low
temperature ($T = 5$ K) current-voltage characteristics between 
the different n-regions (source, drain and collector)
after illumination. We observed a conduction threshold 
between source and drain contacts of $\sim 0.65$ V, 
while the source-gate conduction threshold was found to be 
much higher, owing to the larger distance 
between these electrodes. As expected, $I_{C}$ is negligible 
within the explored range of voltages. 
The observed asymmetry of the I-V curves shown in
Fig.\,\ref{fig1}(c) does not affect the device operation and
was probably due to device inhomogeneities introduced 
during the wet etching process.

SAWs propagating along the (0\={1}\={1}) crystal direction 
were generated by means of an interdigital transducer (IDT)
composed of 100 pairs of 80-$\mu$m-long Al fingers with 1-$\mu$m 
periodicity ($\sim 3$ GHz resonance frequency on
GaAs). Transducers were fabricated at a distance of 500 $\mu$m 
from the n-i-n injector by electron-beam lithography.
The IDT resonance frequency was determined by measuring the 
power reflected by the IDT as a function of the
excitation frequency. The low temperature ($T = 5$ K) frequency 
response displayed a dip at 2.929 GHz with a full width
at half maximum (FWHM) of 2 MHz, consistently with the 
periodicity of the transducer.

\begin{figure}[!ht]
\includegraphics[width=8cm]{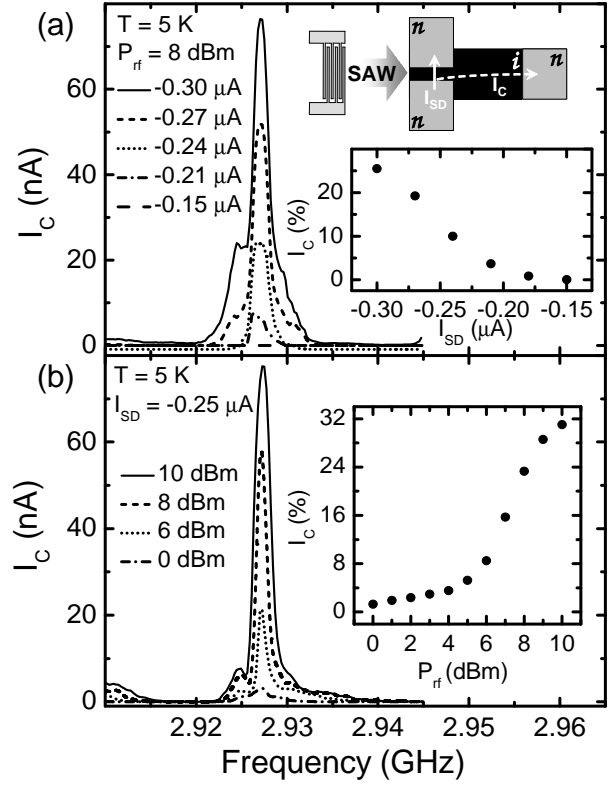}
\caption{(a) Collector current as a function of the frequency 
of the signal applied to the IDT for several values of
the source-drain current at T = 5 K. The radiofrequency power 
was 8 dBm. Inset: measurement scheme and electron
extraction efficiency as a function of the source-drain 
current at T = 5 K. The frequency of signal applied to the IDT
was 2.927 GHz and its power was 8 dBm. (b) Collector current 
as a function of the frequency of the signal applied to
the IDT for several values of the power of the radiofrequency 
signal at T = 5 K and for a fixed value source-drain
current $I_{SD}$ = -0.25 $\mu$A. Inset: electron extraction 
efficiency as a function of the power of the radiofrequency
signal at T = 5 K. The frequency of the signal applied to 
the IDT and the source-drain current were 2.927 GHz and -0.25
$\mu$A respectively.} \label{fig2}
\end{figure}

We monitored $I_{C}$ by means of a 
low-noise current preamplifier in the presence of SAWs while 
injecting a constant source-drain current, $I_{SD}$. The gates 
were left floating and the temperature was set at $T = 5$ K. 
Figure \ref{fig2} (a) shows $I_{C}$ as a function of the 
frequency $f_{rf}$ of the signal applied to the IDT for 
$I_{SD}$ from -0.15 $\mu$A to -0.30 $\mu$A.

A pronounced current peak, corresponding to electrons 
getting to the collector through the intrinsic region, was 
detected at the SAW excitation frequency
for $I_{SD} < -0.15$ $\mu$A. The height of the peak increases
by making $I_{SD}$ more negative. The electron extraction
efficiency, defined as the fraction of $I_{SD}$ detected at 
the collector, also increases [see the inset of Fig.\,\ref{fig2}
(a)], reaching the value of $\sim 25 \%$ for 
$I_{SD} = -0.30 \mu$A. 

To study the SAW extraction efficiency,
$I_{C}$ was measured as a function of the frequency of 
the IDT excitation signal at fixed $I_{SD}$
at different SAW power levels. 
We observed ACT for $P_{rf}\geq 0$ dBm, where 
$P_{rf}$ is the power of the signal applied to the transducer. 
The electron extraction efficiency increases with $P_{RF}$, reaching
approximately the value of $32 \%$ at $P_{rf}=10$ dBm. 

Qualitatively analogue behavior was obtained by fixing the 
n-i-n injector voltage, $V_{SD}$, instead of the current, 
demonstrating the possibility to process both voltage and 
current signals.

By inverting the sign of $V_{SD}$, i.~e. by biasing one lead of the 
n-i-n injector with positive voltage with respect to the ground 
and maintaining the other lead grounded, no ACT was observed. 
Indeed, this corresponds to lowering the conduction band bottom 
of the barrier with respect to the channel. In this regime the
SAW potential is not able to drive electrons from the barrier
to the channel.

All the measurements described above were carried out from $T = 5$ K 
up to room temperature.  The behavior of the device was
essentially determined by the change of the conduction 
threshold of the injector junction. This indeed progressively
increased up to unacceptable values ($>20$ V) 
for T $\sim$120 K. Low conduction threshold was recovered by further
increasing the temperature, but no ACT was observed in this 
high temperature regime.

\begin{figure}[!ht]
\includegraphics[width=8cm]{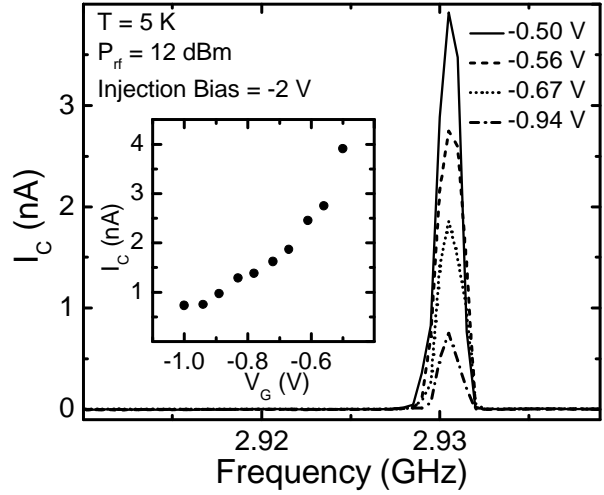}
\caption{Collector current as a function of the frequency 
of the signal applied to the IDT for several values of the
gate voltage at T = 5 K. The radiofrequency power was 12 dBm 
and the n-i-n injector was biased with -2 V. Inset:
collector current as a function of the gate voltage at 
T = 5 K. The frequency and power of signal applied to the IDT
were 2.9305 GHz and 12 dBm respectively. The injection bias 
was -2 V.} \label{fig3}
\end{figure}

Finally, we analyzed the effect of the lateral gates on ACT. 
By applying negative voltages to the gates we expect that the 
channel available for electron transfer to the collector 
become progressively narrower and eventually pinches off. 
The acoustoelectric current toward the OC is thus expected 
to decrease and eventually vanish at sufficiently large 
negative voltage.
Figure \,\ref{fig3} demonstrates the modulation of the ACT 
by means of in-plane lateral gates.
The collector current was measured as a function of the 
frequency of the signal applied to the IDT for several values
of the gate voltage\footnote{Both gates were always biased 
at the same voltage.} and for a fixed injector current
(or bias). The resonance peak was observed to decrease for 
gate voltages more negative than -0.5 V. As shown in the
inset of Fig.\,\ref{fig3}, this effect saturates at $\sim -1$ V. 
Within this range of voltage the peak current was
reduced by $82 \%$. We could not completely suppress ACT because
more negative gate voltages led to current leakage
toward the n-electrodes of the injector. Optimization of the gate-injector geometry can prevent current leakage and allow the use of the gates over a
wider range of voltage.

In conclusion, we demonstrated ACT in an original device 
consisting in a lateral n-i-n injector junction, an etched
intrinsic channel, an output ohmic contact and a pair of 
in-plane gates. We showed that SAWs can extract electrons
from a current flowing through the n-i-n injector and transport 
them toward the output contact. ACT was fully
characterized as a function of the injector current and bias, 
the SAW power and at various temperature from 5 K to room
temperature. Finally, we demonstrated the possibility to 
modulate the acoustoelectric current by means of lateral
in-plane gates. 

This work was supported in part by the European Commission
through the IP SECOQC within FP6 and by MIUR within
FISR "Nanodispositivi ottici a pochi fotoni."

\clearpage


\bibliography{nin_050930}

\clearpage


\end{document}